\documentclass[aps,prd,preprint,showpacs,superscriptaddress]{revtex4}
\usepackage{graphics}
\usepackage{graphicx}
\usepackage{pstricks}
\usepackage{epsfig}

\begin{document}

\title{FCNC in the 3-3-1 model with right-handed neutrinos}

\author{Richard H. Benavides}
\affiliation{Instituto de F\'\i sica, Universidad de Antioquia,
A.A. 1226, Medell\'in, Colombia.}
\affiliation{Instituto Tecnol\'ogico Metropolitano, Facultad de Ciencias, Medell\'in, Colombia}
\author{Yithsbey Giraldo}
\affiliation{Instituto de F\'\i sica, Universidad de Antioquia,
A.A. 1226, Medell\'in, Colombia.}
\affiliation{Departamento de F\'\i sica, Universidad de Nari\~no,
A.A. 1175, Pasto, Colombia.}
\author{William A. Ponce}
\affiliation{Instituto de F\'\i sica, Universidad de Antioquia,
A.A. 1226, Medell\'in, Colombia.}
\begin{abstract}
{Flavor changing neutral currents coming from a new non-universal neutral Gauge Boson and from the non-unitary quark mixing matrix for the $SU(3)_c\otimes SU(3)_L\otimes U(1)_X$ model with right handed neutrinos are studied. By imposing as experimental constraints the measured values of the $3\times 3$ quark mixing matrix, the neutral meson mixing, and bounds and measured values for direct flavor changing neutral current processes, the largest mixing of the known quarks with the exotic ones can be established, with new sources of flavor changing neutral currents being identified. Our main result is that for a $|V_{tb}|$ value smaller than one, large rates of rare top decays such as $t\rightarrow c\gamma$, $t\rightarrow cZ$, and $t\rightarrow cg$ (where $g$ stands for the gluon field) are obtained; but if $|V_{tb}|\sim 1$ the model can survive present experimental limits only if the mass of the new neutral Gauge Bosons becomes larger that 10 TeV.}
\end{abstract}


\maketitle

\section{\label{sec:sec1}Introduction}
The standard model (SM) based on the local gauge group $SU(3)_c\otimes
SU(2)_L\otimes U(1)_Y$ \cite{jf}, with all its successes, fails to explain several fundamental issues such
as: hierarchical charged fermion masses, fermion mixing, charge quantization, strong CP violation, replication of families, neutrino masses and oscillations~\cite{RN}, etc.. All this make us think that we must call for extensions of the model.

The flavor problem encloses two of the most intriguing puzzles in modern particle physics, which are the number of fermion families in nature and the pattern of fermion masses and mixing angles, both in the quark and lepton sectors. With each family being anomaly-free by itself, the SM renders, on theoretical grounds, the number of generations completely unrestricted, except for the indirect bound imposed by the asymptotic freedom of the strong interactions theory, based on the local gauge group $SU(3)_c$, also known as quantum cromo dynamics or QCD.

Many attempts to answer the question of hierarchical quark masses and mixing angles for three families have been reported in the literature, using the top quark as the only heavy quark at the weak scale \cite{hf}. But further insight into the flavor problem can be gained by contemplating the existence of additional heavy quarks.

Popular and well motivated extension of the SM which containt extra heavy quarks are based on the local gauge group~\cite{pf, vl, ozer, sher, pfs, pgs} $SU(3)_c\otimes SU(3)_L\otimes U(1)_X$ (called hereafter 3-3-1 for short). The several possible structures enlarge the SM in its gauge, scalar, and fermion sectors. Let us mention some outstanding features of 3-3-1 models:

\begin{itemize}
\item The simple models are free of gauge anomalies, if and only if the number of families is a multiple of three~\cite{pf, vl, ozer} (becoming just three by imposing QCD asymptotic freedom).
\item A Peccei-Quinn chiral symmetry can be implemented easily~\cite{pq, pal}.
\item One quark family has different quantum numbers than the other two, fact that may be used to explain the heavy top quark mass~\cite{ff, canal}.
\item The scalar sector includes several good candidates for dark matter~\cite{sanchez}. 
\item The lepton content is suitable for explaining some neutrino properties~\cite{kita}. 
\item The hierarchy in the Yukawa coupling constants can be avoided by implementing several universal see-saw mechanisms~\cite{canal, seesaw, dwl}.
\end{itemize}

In the SM with three generations, the quark mixing matrix, called in the literature the Cabibbo-Kobayashi-Maskawa (CKM) mixing matrix~\cite{nc}, is a $3\times 3$ unitary matrix. As a consequence of this unitary character, and for models with only one SM Higgs doublet, the flavor changing neutral currents (FCNC) are absent at tree level, with a strong suppression of the same FCNC at the one-loop level, due to the existence of the Glashow-Iliopoulos-Miani (GIM) mechanism~\cite{sg}. For the minimall 3-3-1 model of Pisano-Pleitez and Frampton~\cite{pf} the quark mixing matrix is the same CKM mixing matrix of the SM, but FCNC at tree level appears due to the existence of a new, non-universal neutral Gauge Boson~\cite{jliw}.

In this analysis we are going to study the FCNC at tree-level and the quark mass spectrum and its mixing matrix, for some 3-3-1 models without exotic electric charges. A classification of all those models has been presented allready in Refs.~\cite{sher, pfs, pgs}. As far as the quark content is concerned, all the three family 3-3-1 models without exotic electric charges fall into four categories: {\bf Category A} which includes models with four up-type quarks and five down-type quarks, {\bf Category B} which includes models with five up-type quarks and four down-type quarks, {\bf Category C} for models with six up-type quarks and three down-type quarks, and {\bf Category D} for models with three up-type quarks and six down-type quarks.

For all the models in the four categories above, the number of up-type quarks is not equal to the number of down-type quarks and thus, the quark mixing matrix looses its unitary character. One outstanding consequence of a nonunitary mixing matrix is the existence of new FCNC processes.

Our aim in this analysis is to see, in the context of some 3-3-1 models without exotic electric charges, how large the mixing between the ordinary and exotic quarks can be, without violating current experimental measurements, both in the $3\times 3$ ordinary quark mixing matrix and in the values and bounds measured for FCNC processes.

This paper is organized as follows: in Sec.~\ref{sec:sec2} we classify in four categories all the 3-3-1 models without exotic electric charges, in Sec.~\ref{sec:sec3} we review the Gauge Boson, the fermion, and the scalar content of the 3-3-1 model with right handed neutrinos, calculate the effective tree-level Hamiltonian for FCNC and introduce the most general quark mass matrices for this model, in Sec.~\ref{sec:sec4} we state the experimental constraints to be respected in the numerical analysis carried through in  Sec.~\ref{sec:sec5}. In Sec.~\ref{sec:sec6} the study of new FCNC processes in the 3-3-1 model with right handed neutrinos is done and in Sec.~\ref{sec:sec7} we present our conclusions. An appendix at the end of the paper justifies the numerical analysis used in the main text.


\section{\label{sec:sec2}3-3-1 models without exotic electric charges} 
In Refs.~\cite{sher, pfs, pgs} the classification of 3-3-1 models without exotic electric charges has been  presented. In this section we will do a short summary of the eight three-family models obtained from the grouping of the following closed sets of fields (closed in the sense that each set includes the antiparticles of each charged particle), where the quantum numbers in parenthesis refer to the $[SU(3)_c,SU(3)_L,U(1)_X]$ representations.

\begin{itemize}
\item $S_1=[(\nu^0_\alpha,\alpha^-,E_\alpha^-);\alpha^+;E_\alpha^+]_L$ with quantum numbers $(1,3,-2/3);(1,1,1)$ and $(1,1,1)$ respectively.
\item $S_2=[(\alpha^-,\nu_\alpha,N_\alpha^0);\alpha^+]_L$ with quantum numbers $(1,3^*,-1/3)$ and $(1,1,1)$ respectively.
\item $S_3=[(d,u,U);u^c;d^c;U^c]_L$ with quantum numbers $(3,3^*,1/3);\; (3^*,1,-2/3);\; (3^*,1,1/3)$ and $(3^*,1,-2/3)$ respectively.
\item $S_4=[(u,d,D);u^c;d^c;D^c]_L$ with quantum numbers $(3,3,0);\; (3^*,1,-2/3);\; (3^*,1,1/3)$ and $(3^*,1,1/3)$ respectively.
\item $S_5=[(e^-,\nu_e,N_1^0);(E^-,N_2^0,N_3^0);(N_4^0, E^+,e^+)]_L$ with quantum numbers $(1,3^*,-1/3)$;$(1,3^*,-1/3)$ and $(1,3^*,2/3)$ respectively.
\item $S_6=[(\nu_e, e^-,E_1^-);(E^+_2,N_1^0,N_2^0);(N_3^0, E^-_2,E_3^-)$;  $e^+; E_1^+; E_3^+]_L$ with quantum numbers $(1,3,-2/3)$; $(1,3,1/3)$; $(1,3,-2/3)$; $(111), (111)$;  and $(111)$ respectively.
\end{itemize}

The former set of fields is exhaustive in the sense that any other set will include either particles with exotic electric charges or 3-3-1 vectorlike representations. The several triangle anomalies for the former six sets are presented in Table I, which in turn allows us to build anomaly-free 3-3-1 models for one, two or more families.

\begin{table}[here!]{{TABLE I: Anomalies for $S_i$}}\label{tabl1}

\begin{tabular}{lcccccc}\hline\hline
Anomalies & $S_1$ & $S_2$ & $S_3$ & $S_4$ & $S_5$ & $S_6$ \\ \hline
$[SU(3)_C]^2U(1)_X$ & 0 & 0 & 0 & 0 & 0 & 0 \\
$[SU(3)_L]^2U(1)_X$ & $-2/3$  & $-1/3$ & 1 & 0& 0 & -1\\
$[Grav]^2U(1)_X$ & 0 & 0 & 0 & 0 & 0 & 0 \\
$[U(1)_X]^3$ & 10/9 & 8/9 & $-12/9$ & $-6/9$& 6/9& 12/9 \\
$[SU(3)_L]^3$ & 1 & $-1$ & $-3$ & 3 & $-3$ & 3\\
\hline\hline
\end{tabular} 
\end{table}

\subsection{\label{sec:sec21}Three family models}
Since data from LEP-I strongly favored  the existence of three families of fermions with light neutrinos, we are going to concentrate in what follows only in models with just three families.

From Table (\ref{tabl1}), only the following eight anomaly free three family models can be constructed:

\begin{itemize}
\item Models in Category A.
\begin{enumerate}
\item[1:] $3S_2+S_3+2S_4$, known in the literature as the 3-3-1 model with right-handed neutrinos~\cite{vl}.
\item[2:] $S_1+S_2+ S_3+ 2S_4+ S_5$, a model without universality in its lepton sector, studied in Ref.~\cite{sher}.
\item[3:] $2S_4 + 2S_5+ S_3+ S_6$.
\end{enumerate}
\item Models in Category B.
\begin{enumerate}
\item[4:] $3S_1+2S_3+S_4$, known in the literature as the 3-3-1 model with exotic electrons~\cite{ozer}.
\item[5:] $S_1+ S_2+ 2S_3+ S_4+ S_6$, a second model without universality in its lepton sector, studied also in Ref.~\cite{sher}.
\item[6:] $S_4 + S_5+ 2S_3+ 2S_6$.
\end{enumerate}
\item Models in Category C.
\begin{enumerate}
\item[7:] $3S_4+ 3S_5$ a three family model, carbon copy of the one family model studied in Ref.~\cite{luis}
\end{enumerate}
\item Models in Category D.
\begin{enumerate}
\item[8:] $3S_3 + 3S_6$ a three family model, carbon copy of the one family model studied in Ref.~\cite{marti}
\end{enumerate}
\end{itemize}

As far as we know, models 3 and 6 above have not been studied in the literature yet.

Due to the fact that the three models in {\bf Category A} have the same quark content (four up type quarks and five down type quarks with the third family of quarks transforming different than the other two), the following  analysis of the FCNC at tree-level and of the quark mass spectrum, is valid for the three models in that Category, including the popular 3-3-1 model with right-handed neutrinos~\cite{vl} (the analysis can be extended in a straightforward way to the other models).


\section{\label{sec:sec3}The 3-3-1 model with right handed neutrinos}
Let us review briefly the so-called 3-3-1 model with right-handed neutrinos:
\subsection{\label{sec:sec31}The Gauge Group} 
As it was stated, the model we are interested in, is based on the local gauge group
$SU(3)_c\otimes SU(3)_L\otimes U(1)_X$ which has 17 gauge bosons: one
gauge field $B^\mu$ associated with $U(1)_X$, the 8 gluon fields $G^\mu$
associated with $SU(3)_c$ which remain massless after spontaneous breaking of the electroweak 
symmetry, and another 8 gauge fields associated with $SU(3)_L$ that we
write for convenience as \cite{pgs}
\begin{equation}\label{maga}
\sum_{\alpha=1}^8\lambda^\alpha A^\mu_\alpha=\sqrt{2}\left(
\begin{array}{ccc}D^\mu_1 & W^{+\mu} & K^{+\mu} \\ W^{-\mu} & D^\mu_2 &
K^{0\mu} \\ K^{-\mu} & \bar{K}^{0\mu} & D^\mu_3 \end{array}\right), 
\end{equation}
where $D^\mu_1=A_3^\mu/\sqrt{2}+A_8^\mu/\sqrt{6},\;
D^\mu_2=-A_3^\mu/\sqrt{2}+A_8^\mu/\sqrt{6}$, and
$D^\mu_3=-2A_8^\mu/\sqrt{6}$. $\lambda_\alpha, \; \alpha=1,2,...,8$, are the eight
Gell-Mann matrices normalized as $Tr(\lambda^\alpha\lambda^\beta)  
=2\delta_{\alpha\beta}$.

The charge operator associated with the unbroken gauge symmetry $U(1)_Q$ 
is given by:
\begin{equation}\label{chargo}
Q=\frac{\lambda_{3L}}{2}+\frac{\lambda_{8L}}{2\sqrt{3}}+XI_3
\end{equation}
where $I_3=Diag.(1,1,1)$ is the diagonal $3\times 3$ unit matrix, and the 
$X$ values are related to the $U(1)_X$ hypercharge and are fixed by 
anomaly cancellation. 
The sine square of the electroweak mixing angle is given by 
\begin{equation}\label{ewk}
S_W^2=3g_1^2/(3g_3^2+4g_1^2)
\end{equation}
where $g_1$ and $g_3$ are the coupling 
constants of $U(1)_X$ and $SU(3)_L$ respectively, and the photon field is 
given by~\cite{vl, pgs} 
\begin{equation}\label{foton}
A_0^\mu=S_WA_3^\mu+C_W\left[\frac{T_W}{\sqrt{3}}A_8^\mu + 
\sqrt{(1-T_W^2/3)}B^\mu\right],
\end{equation}
where $S_W,\; C_W$ and $T_W$ are the sine, cosine and tangent of the electroweak mixing 
angle $\theta_W$, respectively.

There are two weak neutral currents in the model associated with the two neutral weak gauge bosons 
\begin{eqnarray}\nonumber \label{zzs}
Z_0^\mu&=&C_WA_3^\mu-S_W\left[\frac{T_W}{\sqrt{3}}A_8^\mu + 
\sqrt{(1-T_W^2/3)}B^\mu\right], \\ \label{zetas}
Z_0^{\prime\mu}&=&-\sqrt{(1-T_W^2/3)}A_8^\mu+\frac{T_W}{\sqrt{3}}B^\mu,
\end{eqnarray}
and another electrically neutral current associated with the 
gauge boson $K^{0\mu}$. In the former expressions 
$Z^\mu_0$ coincides with the weak neutral current of the SM~\cite{vl, pgs}. The physical fields $Z_1^\mu$ and $Z_2^\mu$ are defined by $Z_1^\mu=\cos\theta Z_0^\mu-\sin\theta Z_0^{\prime\mu}$ and $Z_2^\mu=\sin\theta Z_0^\mu+\cos\theta Z_0^{\prime\mu}$, where $\theta$ is a small mixing angle fixed by phenomenology ($\theta\leq |0.001|$, which in turn implies $M_{Z_2}\geq 2.1$ TeV, with a larger mass bound associated to a smaller mixing angle ~\cite{dwl}).

Using  Eqs. (\ref{foton}) and (\ref{zetas}) we can read the gauge boson $Y^\mu$ 
associated with the $U(1)_Y$ hypercharge of the SM 
\begin{equation} \label{hyper}
Y^\mu=\left[\frac{T_W}{\sqrt{3}}A_8^\mu + 
\sqrt{(1-T_W^2/3)}B^\mu\right].
\end{equation}

Equations (\ref{maga}-\ref{hyper}) presented here are common to all the 3-3-1 gauge structures without exotic electric charges~\cite{vl, ozer, sher} as it is analyzed in Refs.~\cite{pfs, pgs}.

\subsection{\label{sec:sec32}The Fermion sectors}
The quark content for the three families in this model, which is the same for the 3 models in Category A, is the following:
$Q^i_{L}=(u^i,d^i,D^i)_L\sim(3,3,0)$, $i=1,2$ for two families, where
$D^i_L$ are two extra quarks of electric charge $-1/3$; $Q^3_{L}=(d^3,u^3,U)_L\sim (3,3^*,1/3)$, where
$U_L$ is an extra quark of electric charge 2/3. The right handed quarks which belong to $SU(3)_L$ singlets 
are $u^{ac}_{L}\sim (3^*,1,-2/3)$, $d^{ac}_{L}\sim (3^*,1,1/3)$ with
$a=1,2,3$ a family index, $D^{ic}_{L}\sim (3^*,1,1/3)$, $i=1,2$, and
$U^c_L\sim (3^*,1,-2/3)$.

The lepton content is given by the three $SU(3)_L$ triplets $L_{lL} =
(l^-,\nu_l^0,\nu_l^{0c})_L\sim (1,3^*,-1/3)$, for $l=e,\mu,\tau$ 
a lepton family index, and the three singlets $l^+_{L}\sim(1,1,1)$, 
where $\nu_l^0$ is the neutrino field associated with the lepton $l^-$, and $\nu_l^{0c}$ plays the role of the right-handed neutrino field associated to the same flavor. For this model universality for the known leptons in the three families is present at tree level in the weak basis.

\subsection{\label{sec:sec33}The scalar sector}
The following is the set of scalar fields and Vacuum Expectation Values (VEV) used in order to break the symmetry and to give a consistent mass spectrum to the fermion fields~\cite{vl}:
\begin{eqnarray}\label{higgs}
\langle\phi_1^T\rangle &=&\langle(\phi^+_1, \phi^0_1,\phi^{'0}_1)\rangle =
\langle(0,0,V)\rangle \sim (1,3,1/3); \\  \nonumber
\langle\phi_2^T\rangle &=&\langle(\phi^+_2, \phi^0_2,\phi^{'0}_2)\rangle =
\langle(0,v_1,0)\rangle \sim (1,3,1/3); \\ \nonumber
\langle\phi_3^T\rangle &=&\langle(\phi^0_3, \phi^-_3,\phi^{'-}_3)\rangle =
\langle(v_2,0,0)\rangle \sim (1,3,-2/3); \\ \nonumber
\end{eqnarray}
with the hierarchy $v_1\sim v_2\sim 10^2$ GeV $<< V\sim$ TeV.

The analysis shows that this set of VEV breaks the
$SU(3)_c\otimes SU(3)_L\otimes U(1)_X$ symmetry in two steps following the scheme

\begin{eqnarray*}
{\mbox 3-3-1}&\stackrel{V}{\longrightarrow}&
SU(3)_c\otimes SU(2)_L\otimes U(1)_Y \\
&\stackrel{v_i}{\longrightarrow}& SU(3)_c\otimes U(1)_{EM},
\end{eqnarray*}
for $i=1,2$ and $U(1)_{EM}$ the Abelian gauge group of the electromagnetism.

\subsection{\label{sec:sec34}FCNC at tree level}
In the context of most of the 3-3-1 models considered in this paper, the third family of quarks is treated differently than the other two; so, it has different couplings to the scalars as well as to the new neutral current $J^\mu_{Z^\prime}$ present in the model (the quark couplings to the SM neutral current $J^\mu_{Z}$ is not only diagonal in flavor but also it is universal). Due to this, new FCNC at tree level  show up, which in principle contribute to FCNC processes which are severely constrained by experiment, most notably by meson mixing~\cite{jliw}.

For the 3-3-1 model with right-handed neutrinos, all the currents were allready calculated in Ref.~\cite{vl}. Using for the photon field $A_\mu$ the expression in Eq. (\ref{foton}) and for $Z_\mu$ and $Z_\mu^\prime$ the definitions in (\ref{zetas}),  the neutral currents, associated with the Hamiltonian
\begin{equation}
 H^0=eA^\mu J_\mu (EM) + (g_3/C_W)Z^\mu J_\mu (Z)+(g_1/\sqrt{3})Z^{'\mu}J_\mu (Z'),
\end{equation}
are 
\begin{eqnarray}
\nonumber J_\mu(EM)&=&\frac{2}{3}(\sum_{a=1}^3\bar{u}_{a}\gamma_\mu u_{a} + \bar{U}\gamma_\mu U\\ \nonumber && -\frac{1}{3}(\sum_{a=1}^3\bar{d}_{a}\gamma_\mu d_{a}+\sum_{i=1}^2\bar{D}_{i}\gamma_\mu D_{i})\\ \nonumber
&& -\sum_{l=e,\mu,\tau} \bar{l}^-\gamma_\mu l\\ 
J^\mu (EM) &=& \sum_f q_f\bar{f}\gamma^\mu f,\\ \nonumber
J^\mu (Z) &=& J^{\mu}_L(Z)-S^2_WJ^\mu (EM),\\ \nonumber
J^\mu (Z') &=& T_WJ^\mu (EM)-J^{\mu}_L(Z'), 
\end{eqnarray}
where $e=gS_W=g'C_W\sqrt{1-T^2_W/3}>0$ is the electric charge, $q_f$ is the electric charge of the fermion $f$ in units of $e$, 
$J^\mu(EM)$ is the electromagnetic current, and the left-handed currents are given by

\begin{eqnarray}\nonumber
J_{L}^\mu (Z)&=&\frac{1}{2}[\sum_{a=1}^3(\bar{u}_L^a\gamma^\mu u_{L}^a-\bar{d}_L^a\gamma^\mu d_L^a)\\ \nonumber 
&+&\sum_l(\bar{\nu}_{lL}\gamma^\mu \nu_{lL}-\bar{l}_L^-\gamma^\mu l_L^-)]\\ \label{ncsm}
&=&\sum_f \bar{f}_LT_{3f}\gamma^\mu f_L,
\end{eqnarray}
and
\begin{eqnarray}\nonumber 
J^{\mu}_L(Z') &=& S_{2W}^{-1}(\bar{u}_{1L}\gamma^\mu u_{1L}+\bar{u}_{2L}\gamma^\mu u_{2L}-\bar{d}_{3L}\gamma^\mu d_{3L}\\ \nonumber 
&&-\sum_l\bar{l}_{l}^-\gamma^\mu l_L^-) \\ \nonumber 
&& + T_{2W}^{-1}(\bar{d}_{1L}\gamma^\mu d_{1L}+\bar{d}_{2L}\gamma^\mu d_{2l}-\bar{u}_{3L}\gamma^\mu u_{3L}\\ \nonumber
&&-\sum_l \bar{\nu}_{lL}\gamma^\mu \nu_{lL})\\ \nonumber
&& + T_{W}^{-1}(\bar{D}_{1L}\gamma^\mu D_{1L} + \bar{D}_{2L}\gamma^\mu D_{2L}- \bar{U}_{L}\gamma^\mu U_{L} \\ \label{ncex}
&&-\sum_l \bar{\nu}_{lL}^{oc}\gamma^\mu \nu_{lL}^{oc})\equiv \sum_f\bar{f}_LT_{3f}'\gamma^\mu f_L,
\end{eqnarray}
with $T_{3f}=diag(1/2, -1/2,0). \;\; T_{3f}'=diag(S_{2W}^{-1},T_{2W}^{-1},-T_{W}^{-1})$ is a convenient $3\times 3$ diagonal matrix (both marices $T_{3f}$ and $T_{3f}^\prime$ acting on the representation 3 of $SU(3)_L$, with their negative values when acting on the representation $3^*$). $f$ is a generic symbol for the representation 3 (and 3*) of $SU(3)_L$\cite{vl}, and $J^\mu_L(Z^\prime)$ allthough diagonal in the weak basis is not universal.

The couplings of the left-handed quarks with the $Z^\prime$ Gauge Boson, can then be written in the form
\begin{equation}
\mathcal{L}(Z^\prime)=\frac{e}{\sqrt{3-4S^2_W}}Z^{'\mu}J_{\mu}(Z^\prime),
\end{equation}
with 
\begin{eqnarray}
J^\mu(Z^\prime)=\frac{1}{S_{2W}}\sum_f\bar{f}\gamma^\mu [S^2_W Y -2\sqrt{3}C^2_W T_{8L}]P_Lf,
\end{eqnarray}
where $P_L=(1-\gamma^5)/2$. Since the value of $T_{8L}$ is different for triplets and antitriplets, the $Z^\prime$ coupling is different for the third family and we have FCNC a tree level. These currents can be written in the form:
\begin{equation}
J_{Z^\prime(FCNC)}^{\mu}=-\frac{\sqrt{3}}{T_W}\sum_f\bar{f}\gamma^{\mu}[T_{8L}-T_{8L}^*]P_L f=\frac{1}{T_W}\sum_f\bar{f}\gamma^{\mu}P_L f,
\end{equation}
with the tree level effective Lagrangian for these FCNC calculated to be
\begin{equation}
\mathcal{L}_{(FCNC)}= \frac{g_3 C_W}{\sqrt{(3-4S^2_W)}}(S_\theta Z_1^\mu + C_\theta Z_2^\mu)\sum_f \bar{f}\gamma^\mu P_L f,
\end{equation}
where $\theta$ is the mixing angle between the two massive neutral Gauge Bosons $Z$ and $Z^\prime$ which defines the physical states $Z_1$ and $Z_2$ respectively (this angle is very small as can be seen from the last paper in Ref.~\cite{vl}).

Beacause the third family of quarks is treated differently we have that 

\begin{widetext}
\begin{equation}\label{jzp}
J_{Z'}^{\mu}=[\bar{\vec \mathcal{U}}\gamma^{\mu}P_LV_L^{u\dag}\left(
\begin{array}{cccc}S_{2W}^{-1} &  & & \\ & S_{2W}^{-1} & & \\ &  & -T_{2W}^{-1} & \\ &  &  &  -T_W^{-1} \end{array}\right)V^u_L{\vec{\mathcal{U}}} + \bar{\vec{\mathcal{D}}}\gamma^{\mu}P_LV_L^{d\dag}\left(
\begin{array}{ccccc}T_{2W}^{-1} &  & & & \\ & T_{2W}^{-1} & & & \\ &  & -S_{2W}^{-1} & & \\& & & T_W^{-1} & \\ & & & & T_{W}^{-1} \end{array}\right)V^d_L{\vec{\mathcal{D}}}],
\end{equation}
\end{widetext}
where ${\vec\mathcal{U}}$ and ${\vec\mathcal{D}}$ are four column and five column vectors for the up and down quark sectors respectively, and $V^u_L$ and $V^d_L$ are the $4\times 4$ and $5\times 5$ unitary matrices which diagonalize the mass matrices of the up and down quark sectors respectively, with $V_{mix}=V^u_LV_L^{d\dagger}$ the non-unitary $4\times 5$ quark mixing matrix in the context of this particular model (see the following Section). As can be seen, $J^\mu_{Z^\prime}$ in Eq.~(\ref{jzp}) induced FCNC a tree level.

Using the tree-level current in Eq.~(\ref{jzp}), the following effective Hamiltonian can be obtained
\begin{equation}\label{Heff}
|\mathcal{H}_{eff}|^2=\frac{4\sqrt{2}G_F C^4_W C^2_\theta }{(3-4S^2_W)}|V_{Jj\alpha}^*V_{Lj\beta}|^2 
(\frac{M_{Z_1}^2}{M_{Z_2}^2}+T_\theta ^2)(\alpha_L\gamma^\mu \beta)^2,
\end{equation}
which can be used to calculate the tree-level diagrams for $K^0-\bar{K^0},\; D^0-\bar{D}^0,\; B_d^0-\bar{B}_d^0$ and $B_s^0-\bar{B}_s^0$ mixing just by replacing  $(\alpha, \beta)$ by $(d,s),\; (u,c),\; (d,b)$ and $(s,b)$ respectively. An equation similar to (\ref{Heff}) but for the minimall model~\cite{pf} has been derived in Ref.~\cite{Liu}.

\subsection{\label{sec:sec35}Mass matrices}
In this subsection we are going to present the most general quark mass matrices for all the 3-3-1 three family models without exotic electric charges belonging to Category A, and to set our notation.

The Higgs scalars introduced above are used to write the Yukawa terms for the quarks. In the case of the up quark sector, the most
general invariant Yukawa Lagrangian is given by
\begin{eqnarray}\label{mup}
{\cal L}^u_Y&=&
\sum_{\alpha=1,2}Q_L^3\phi_\alpha C(h^U_\alpha
U_L^c+\sum_{a=1}^3h_{a\alpha}^uu_L^{ac}) \\ \nonumber
&& + \sum_{i=1}^2Q^i_L\phi_3^*
C(\sum_{a=1}^3h^{u\prime}_{ia}u_L^{ac}+h_i^{U\prime}U_L^c)
+ h.c.,
\end{eqnarray}
where $C$ is the charge conjugation operator. In the weak basis ${\cal \vec{U}}=(u^1,u^2,u^3,U)$ the former Lagrangian produces the following $4\times 4$ quark mass matrix for the up quark sector
\begin{equation}\label{maup}
M_U=\left(\begin{array}{cccc} 
v_2h_{11}^{u\prime} & v_2h_{12}^{u\prime} & v_2h_{13}^{u\prime} & v_2h_1^{U\prime} \\ 
v_2h_{21}^{u\prime} & v_2h_{22}^{u\prime} & v_2h_{23}^{u\prime} & v_2h_2^{U\prime} \\ 
v_1h_{12}^{u} & v_1h_{22}^{u} & v_1h_{32}^u & v_1h_2^U \\
Vh_{11}^u & Vh_{21}^u & Vh_{31}^u & Vh_1^U \\ 
\end{array}\right).
\end{equation}

For the down quark sector, the most general Yukawa Lagrangian is now 

\begin{eqnarray}\label{mdown} \nonumber
{\cal L}^d_Y &=& \sum_{\alpha=1,2}
\sum_{i}Q^i_L\phi_\alpha^*C(\sum_ah^d_{ia\alpha}d_L^{ac}
+\sum_jh^D_{ij\alpha}D_L^{jc}) \\
&&+ Q_L^3\phi_3C(\sum_ih^D_iD_L^{ic}+\sum_{a}h_a^dd_L^{ac})+h.c..
\end{eqnarray}
which in the weak basis ${\cal \vec{D}}=(d^1,d^2,d^3,D^1,D^2)$ produces the following $5\times 5$ quark mass matrix for the down quark sector

\begin{equation}\label{madown}
M_D=\left(\begin{array}{ccccc} 
v_1h_{112}^d & v_1h_{122}^d & v_1h_{132}^d & v_1h_{112}^D & v_1h_{122}^D \\
v_1h_{212}^d & v_1h_{222}^d & v_1h_{232}^d & v_1h_{212}^D & v_1h_{222}^D \\
v_2h_1^d & v_2h_2^d & v_2h_3^d & v_2h_1^D & v_2h_2^D \\
Vh_{111}^d & Vh_{121}^d & Vh_{131}^d & Vh_{111}^D & Vh_{121}^D \\
Vh_{211}^d & Vh_{221}^d & Vh_{231}^d & Vh_{211}^D & Vh_{221}^D \\
\end{array}\right).
\end{equation}
$M_U$ and $M_D$ in (\ref{maup}) and (\ref{madown}) must be diagonalized in order to get the mass eigenstates which exist in nature, defining in this way a non-unitary $4\times 5$ quark mixing matrix of the form 

\begin{equation}\label{dmix}
V_{mix}\equiv V_L^u{\cal P}V_L^{d\dag}=\left(\begin{array}{ccccc} 
V_{ud} & V_{us} & V_{ub} & V_{ub^\prime} &V_{ub^{\prime\prime}} \\ 
V_{cd} & V_{cs} & V_{cb} & V_{cb^\prime} & V_{cb^{\prime\prime}}\\
V_{td} & V_{ts} & V_{tb} & V_{tb^\prime} & V_{tb^{\prime\prime}}\\
V_{t^\prime d} & V_{t^\prime s} & V_{t^\prime b} & V_{t^\prime b^\prime}& V_{t^\prime b^{\prime\prime}}\\
\end{array}\right),
\end{equation}
where $V_L^u$ and $V_L^d$ are $4\times 4$ and $5\times 5$ unitary matrices which diagonalize $M_UM_U^\dag$ and $M_DM_D^\dag$ respectively, and ${\cal P}$ is the projection matrix over the ordinary quark sector (in the weak basis, the exotic quarks transform as singlets under $SU(2)_L$ transformations, thus they do not couple with the $W^\pm$ Gauge Bosons). This matrix is given by
\begin{equation}\label{calp}
{\cal P}=\left(\begin{array}{ccccc} 
1 & 0 & 0 & 0 & 0 \\
0 & 1 & 0 & 0 & 0 \\
0 & 0 & 1 & 0 & 0 \\
0 & 0 & 0 & 0 & 0 \\ 
               \end{array}\right).
\end{equation}
$V_{mix}$ in (\ref{dmix}) defines the couplings of the physical quark states.
$(u,c,t,t^\prime)$ and $(d,s,b,b^\prime,b^{\prime\prime})$ with the charged current associated with the weak gauge boson $W^+$.


\section{\label{sec:sec4}Experimental Constraints.}
In the quark sector, several parameters have been measured with high accuracy, with values which constitute some of the strongest  experimental constraints for model builders. The following three sets of numbers are going to be considered in what follows:

\subsection{\label{sec:sec41}The $3\times 3$ quark mixing matrix}
The masses and mixing of quarks in the SM come from Yukawa interaction terms with the Higgs condensate, which produces two $3\times 3$ quark mass matrices for the up and down quark sectors; matrices that must be diagonalized in order to identify the mass eigenstates. The unitary CKM quark mixing matrix ($V_{CKM}\equiv V^u_{3L}V^{d\dag}_{3L}$) couples the six physical quarks to the charged weak gauge boson $W^+$, where $V_{3L}^u$ and $V_{3L}^d$ are now the diagonalizing unitary $3\times 3$ matrices of the SM up and down quark sectors respectively.

The unitary matrix $V_{CKM}$ has been parametrized in the literature in several different ways, but the most important fact related with this matrix is that most of its entries have been measured with high accuracy, with the following experimental limits~\cite{pdg}:
\begin{widetext}
\begin{equation}\label{maexp}
 V_{exp}=
\left(\begin{array}{ccc}
0$.$970\leq  |V_{ud}|\leq 0$.$976 & 0$.$223\leq |V_{us}|\leq 0$.$228 & 0$.$003\leq |V_{ub}|\leq0$.$005\\
0$.$219\leq |V_{cd}| \leq 0$.$241 & 0$.$90\leq |V_{cs}| \leq 1$.$0 & 0$.$039 \leq |V_{cb}| \leq 0$.$045\\
0$.$006\leq |V_{td}| \leq 0$.$008 & 0$.$036\leq |V_{ts}| \leq 0$.$044 & |V_{tb}| \geq0$.$78
\end{array}\right).
\end{equation}
\end{widetext}

The numbers quoted in matrix (\ref{maexp}), which are measured at the Fermi scale $(\mu~\approx M_Z)$~\cite{koide}, are generous in the sense that they are related to the direct experimental measured values, some of them at 90\% coffidence level, with the largest uncertainties taken into account, without bounding the numbers to the orthonormal  constrains on the rows and columns of a $3\times 3$ unitary matrix. In this way we leave the largest room available for possible new physics, respecting the well measured values in $V_{exp}$.

The most conservative alternative of using numerical entries which take into account unitary constraints in $V_{exp}$ is going to be considered also at the end of our study.

\subsection{\label{sec:sec42}Direct FCNC searches}
The unitary character of the SM mixing matrix $V_{CKM}$ implies flavor diagonal couplings of all the neutral bosons of the SM (such as Z boson, Higgs boson, gluons and photon) to a pair of quarks, giving as a consequence that no FCNC are present at tree level. At one-loop level, the charged currents generate FCNC transitions via penguin and box diagrams~\cite{jf}, but they are highly suppressed by the GIM mechanism~\cite{sg}. For example, FCNC processes in the charm  sector $(c\rightarrow u\gamma)$ were calculated in the context of the SM in Ref.~\cite{GB}, giving a branching ratio suppressed by 15 orders of magnitude, leaving in this way a large window of opportunities for new physics in charm decays.

To date, the following direct FCNC branching ratios and bounds have been measured in several experiments:
\begin{itemize}
\item $\mathcal{B}r[b\rightarrow s\gamma]=(3.52\pm 0.24)\times 10^{-4}$ ~\cite{hfag}
\item $\mathcal{B}r[B\rightarrow K^*l^+l^-]= (1.68\pm 0.86)\times 10^{-6}$ ~\cite{BB},
\item $\mathcal{B}r[s\rightarrow d\gamma(dl^+l^-)]<10^{-8}$ ~\cite{sp}
\item $\mathcal{B}r[c\rightarrow ul^+l^-]<4\times 10^{-6}$ ~\cite{vm}
\item $\mathcal{B}r[b\rightarrow sl^+l^-,dl^+l^-]<5\times 10^{-7}$ ~\cite{VM},
\end{itemize}
with $l=e,\mu$. In our study, these ratios and bounds are also going to be respected. Important to mention here that the SM next to next to leading order calculation for $\mathcal{B}r[b\rightarrow s\gamma]$ is $(3.60\pm 0.30)\times 10^{-4}$~\cite{gambi}, allready in agreement with the measured value, which   constitutes a very sensitive prove of new physics.

\subsection{\label{sec:sec43}Indirect FCNC searches}
In general, flavor physics processes and in particular meson mixing, are known to constraint FCNC of the type produced by a non-universal $Z^\prime$ Gauge Boson. At present the most severe constraints arise from $K^0,\; D^0,\; B^0_d$ and $B^0_s$ neutral meson mixing. To date, the following experimental measurements have been obtained~\cite{pdg}:
\begin{itemize}
\item $\Delta m_{K^0}=0.5290\pm 0.0016\times 10^{10}\hbar s^{-1}$
\item $\Delta m_{D^0}=7\times 10^{10}\hbar s^{-1}$
\item $\Delta m_{B_d^0}=0.507\pm 0.005\times ps^{-1}$~\cite{abul}.
\item $\Delta m_{B_s^0}=17.77\pm 0.17\;\; ps^{-1}$~\cite{abul},
\end{itemize}
numbers which severely constraint models with FCNC occurring at the tree-level.

\section{\label{sec:sec5}Numerical Analysis.}
As it is expected from Eq.~(\ref{Heff}), FCNC at tree-level are depleted when the ordinary quarks mix with the exotic ones, the largest the mixing, the smaller the FCNC effects. 
In this section we are going to see, in the context of the 3-3-1 model with right-handed neutrinos, how large the quark mixing can be,  without violating the experimental measured values quoted in the previous section.

In the analysis we assume that $v_1=v_2\equiv v=123$ GeV, value  supported by the result $M_W^2=g_3^2(v_1^2+v_2^2)/2$~\cite{vl}  with $g_3$ the gauge coupling constant of $SU(3)_L$ (that is equal to $g_2$, the gauge coupling constant of $SU(2)_L$ in the SM), and also we use $V=1$ TeV, the 3-3-1 mass scale which fixes the mass values for all the new fermions of the different models.

\subsection{\label{sec:sec51}The $4\times 5$ mixing matrix}
In this section we are going to study the non-unitary $4\times 5$ quark mixing matrix $V_{mix}$ in Eq.~(\ref{dmix}) for the three models in category A (models with four up-type quarks and five down-type quarks) including the 3-3-1 model with right handed neutrinos. What we pretend to do is to look for the maximall mixing of the ordinary quarks with the exotic ones, without violating the experimental constraints quoted in the previous section.

Let us start first with what we have called the down-up approach, which consists of looking for quark mass matrices which fit the experimental constraints of $V_{exp}$ in (\ref{maexp}), with a value $V_{tb}\sim 0.8$, the smallest possible. The numerical analysis suggest to start with the following orthogonal quark mass matrices 
\begin{equation}\label{upte4}
M^u_4 =\left(\begin{array}{cccc} 
0.00047 & 0.02812 & 0 & 0 \\ 
0.02812 & 0.580 & 0 & 0  \\
0 &  0 & 171.7  & 0\\
0 & 0 & 0 & m_{t^\prime} \\
\end{array}\right)
\end{equation}

\begin{equation}\label{dowte5}
M^d_5 =\left(\begin{array}{ccccc} 
0.018 & -0.4288 & -2.63 & -3.41 & 0\\ 
-0.4288 & 9.316 & 57.608 & 75.98 & 0 \\
-2.63 & 57.608 & 361.8 & 472.4 & 0 \\
-3.41 & 75.98 & 472.4 & 624.5 & 0 \\
0 & 0 & 0 & 0 & m_{b^{\prime\prime}} \\
\end{array}\right),
\end{equation}
which for $m_{t^\prime}=m_{b^{\prime\prime}}=1500$ GeV, reproduce the following set of eigenvalues (in units of GeV)
\[m_t=171.7,\; m_c=0.582, \; m_u=1.4\times 10^{-3}\]
\[m_b=2.83,\; m_s=0.069, \; m_d=3.4\times 10^{-3};\]
\[m_{t^\prime}=1500,\; m_{b^{\prime\prime}}=1500, \; m_{b^\prime}=993, \]
numbers to be compared with the values quoted in the appendix (taken from the second paper in Ref.~\cite{koide}).

The rotation matrices which diagonalize $M^u_4$ and $M^d_5$ are
\begin{equation}\label{rot4}
V_4^u=\left(
\begin{array}{cccc}
0.9984 & -0.0563 & 0 & 0 \\
0.0563 & 0.9984 & 0 & 0\\
0 & 0 & 1 & 0 \\
0 & 0 & 0 & 1 \end{array}\right)_{ru},
\end{equation}
and
\begin{equation}\label{rot5}
V_{5}^d=\left(
\begin{array}{ccccc}
0.9850 & 0.172 & 0.006 & -0.02 & 0 \\
0.1724 & -0.9798 & 0.031 & 0.097 & 0 \\
0.011  & 0.0366 & -0.798 & 0.602 & 0 \\
-0.0044 & 0.0965 & 0.602 & 0.7925 & 0 \\
0 & 0 & 0 & 0 & 1 \\ \end{array}
\right)_{rd}.
\end{equation}

Matrices which combine to produce the folowing non-unitary $4\times 5$ mixing matrix $V_{mix}^{4\times 5}=|V_4^u{\cal P}V_5^{d\dagger}|$
\begin{equation}\label{mix45}
V_{mix}^{4\times 5} =\left(\begin{array}{ccccc} 
0.974 & 0.227 & 0.008 & 0.0098 & 0 \\ 
0.227 & 0.9685 & 0.0371 & 0.096 & 0 \\
0.0060 &  0.031  & 0.798 & 0.602 & 0 \\
0 & 0 & 0 & 0 & 0 \\
\end{array}\right),
\end{equation}
numbers to be compared with the experimental limits in (\ref{maexp}) and with the numbers quoted in the appendix for $V_{mix}^{ud}$ in (\ref{rotu4}) for the up-down approach.


\section{\label{sec:sec6} New FCNC processes}
Next, we are going to evaluate the new contributions to the FCNC processes coming from the nonunitary character of $V^{4\times 5}_{mix}$ in Eq.~(\ref{mix45}), and from the rotation matrices $V^4_u$ and $V^d_5$.
\subsection{Penguin processes for the SM quarks}
The following are the penguin contributions to the FCNC coming from $V^{4\times 5}_{mix}$:
\subsubsection{The bottom sector}
Let us evaluate first the electromagnetic penguin contribution to ${\cal B}r^{t}(b\rightarrow s\gamma)$ coming from the $t$ quark, calculated with the expectator model, scaled to the semileptonic decay $b\rightarrow q_il\nu_l,\; q_i=c,u$, and without including QCD corrections (which are small for the $b$ sector~\cite{jf}). This value is calculated to be~\cite{GB}

\begin{equation}\label{fincal}
{\cal B}r^{t}(b\rightarrow s\gamma)\approx\frac{3\alpha}{2\pi}
\frac{|V_{tb}^*V_{ts}F^Q(x^{t})|^2}{[f(x_c)|V_{cb}|^2+f(x_u)|V_{ub}|^2]}B_{B\rightarrow Xl\nu_l},
\end{equation}
where $\alpha$ is the fine structure constant, $B_{B\rightarrow Xl\nu_l}\approx 0.1$ is the branching ratio for semileptonic $b$ meson decays taken from Ref.~\cite{pdg}, $x^{t}=(m_{t}/M_W)^2$, $x_c=m_c/m_b$ and $x_u=m_u/m_b$. $F^Q(x)$ is the contribution of the internal heavy quark line to the electromagnetic penguin given by

\begin{eqnarray*}
F^Q(x)&=&Q\left[\frac{x^3-5x^2-2x}{4(x-1)^3}+\frac{3x^2\ln{x}}{2(x-1)^4}\right] \\
&+&\frac{2x^3+5x^2-x}{4(x-1)^3}-\frac{3x^3\ln{x}}{2(x-1)^4},
\end{eqnarray*}
where $Q=2/3$ for $t$ in the quark propagator [$Q=-1/3$ and $x=x^{b^\prime}=(m_{b^\prime}/M_W)^2$ when $b^\prime$ propagates, with the appropriate changes when $b^{\prime\prime}$ propagates] and $f(x_i)$ is the usual phase space factor in semileptonic meson decay, given by~\cite{jf}

\[f(x)=1-8x^2+8x^6-x^8-24x^4\ln{x}.\]

For the numerical evaluations of ${\cal B}r^{t}(b\rightarrow s\gamma)$, let us use the values $\alpha(1GeV)=1/135$, $m_{t}=171.7$ GeV, $m_c=0.6$ GeV, $m_b=2.8$ GeV and $m_u=1.4$ MeV~\cite{koide} (which are not the pole values). Using these numbers we obtain: $F^{2/3}(x^{t})\approx 0.387$, $f(x_c)\approx 0.72$ and $f(x_u)\approx 1$. Plug in the numbers in Eq.~(\ref{fincal}) and using the values for $V_{mix}^{4\times 5}$ in equation (\ref{mix45}) for the couplings of the physical quark states, we get

\[{\cal B}r^{t}(b\rightarrow s\gamma)\approx 3\times 10^{-5},\]
close to the SM calculation as it should be, since this process does not receive a contribution from the exotic quarks.

The former analysis can be used also to estimate the branching ratios for the rare gluon penguin decay $b\longrightarrow sg$, where $g$ stands for the gluon field. The results is
\begin{eqnarray*}
{\cal B}r^{t}(b\rightarrow sg)&=&\frac{\alpha_s(1GeV)}{\alpha(1GeV)}{\cal B}r^{t}(b\rightarrow s\gamma)\\ 
&\approx&13{\cal B}r^{t}(b\rightarrow s\gamma)\approx 3.9\times 10^{-4},
\end{eqnarray*}
a process difficult to meassure due to the hadronization of the gluon field $g$. (This last process is of the same order of magnitude of the virtual weak penguin bottom process $b\longrightarrow sZ$).

A similar analysis shows that 
\[{\cal B}r^{t}(b\rightarrow d\gamma)=\frac{|V_{td}|^2}{|V_{ts}|^2}{\cal B}r^{t}(b\rightarrow s\gamma)\approx 1.16\times 10^{-6},\]
which is safe and in agreement with the bounds quoted in Section (\ref{sec:sec42}).

\subsubsection{The strange sector}
In a similar way we can evaluate ${\cal B}r^{t}(s\rightarrow d\gamma)$ scaled to the semileptonic decay $s\rightarrow ul\nu_l$, which is given now by 
\begin{equation}\label{fincs}
{\cal B}r^{t}(s\rightarrow d\gamma)\approx\frac{3\alpha}{2\pi}
\frac{|V_{ts}^*V_{td}F^{2/3}(x^{t})|^2}{f(x^\prime_u)|V_{us}|^2}B_{K\rightarrow \pi l\nu_l}.
\end{equation}

With $x^\prime_u=m_u/m_s,\; m_s$(1GeV)=69 MeV, and $B_{K\rightarrow \pi l\nu_l}\approx 5\times 10^{-2}$ taken from Ref.~\cite{pdg}, we get 
\[{\cal B}r^{t}(s\rightarrow d\gamma)\approx 1.75\times 10^{-11},\]
in agreement with the experimental bound quoted in Section (\ref{sec:sec42}).

\subsubsection{The charm sector}
Now let us evaluate ${\cal B}r^{b^\prime}(c\rightarrow u\gamma)$ scaled to the semileptonic decay $c\rightarrow q_jl\nu_l$, where $q_j=s,d$. The branching ratio is

\begin{equation}\label{finccu}
\frac{{\cal B}r^{b^\prime}(c\rightarrow u\gamma)}{B_{D\rightarrow X_s l\nu_l}}\approx\frac{3\alpha}{2\pi}
\frac{|(V_{cb^\prime}^*V_{ub^\prime}) F^{-1/3}(x^{b^\prime})|^2}{[f(x_s)|V_{cs}|^2+f(x_d)|V_{cd}|^2]},
\end{equation}
where $x_s=m_s/m_c,\; x_d=m_d/m_c$. With $B_{D\rightarrow X_s l\nu_l}\approx 0.2$ taken from Ref.~\cite{pdg}, $F^{-1/3}(x^{b^\prime})\approx 0.3849$, $f(x_s)\approx 0.895$ for $m_s=150$ MeV and $f(x_d)\approx 1$, for $m_d=3.4$ MeV, we get 
\[{\cal B}r^{b^\prime}(c\rightarrow u\gamma)\approx 1. \times 10^{-10},\]
five orders of magnitude larger than the SM prediction~\cite{GB}, but still unobservable small. Of course, the quantum QCD corrections for this decay could be quite large (see the second paper in Ref.~\cite{GB}).

\subsubsection{The top sector}
We proceed this analysis with the study of the FCNC for the top quark in the context of the three 3-3-1 models in category A. As we are about to see, some of the predictions are ready to be tested at the Large Hadron Collider (LHC).

In the SM, the one-loop induced FCNC for the top quark have a strong GIM suppression, resulting in negligible branching ratios for top FCNC decays. The SM values predicted are~\cite{saav}: ${\cal B}r^{SM}(t\rightarrow c\gamma)\approx 4.6\times 10^{-14}$, and 
${\cal B}r^{SM}(t\rightarrow cg)\approx 4.6\times 10^{-12}$.

The new FCNC ${\cal B}r^{b^\prime}(t\rightarrow c\gamma)$ and ${\cal B}r^{b^\prime}(t\rightarrow u\gamma)$ predicted for the top quark in the context of the 3-3-1 model with right-handed neutrinos, scaled to the semileptonic decay $t\rightarrow q_k l\nu_l,\; q_k=b,s,d$; are given by 

\begin{equation}\label{finct}
\frac{{\cal B}r^{b^\prime}(t\rightarrow c\gamma)}{B_{T\rightarrow Xl\nu_l}}\approx\frac{3\alpha}{2\pi}
\frac{|(V_{tb^\prime}^*V_{cb^\prime}) F^{-1/3}(x^{b^\prime})|^2}{[f(x_b)|V_{tb}|^2+f(x_s)|V_{ts}|^2]}
\end{equation}
which we evaluate at the $m_t=171.7$ GeV, the pole mass scale for the top quark, which gives

\[{\cal B}r^{b^\prime}(t\rightarrow c\gamma)\approx 2.75\times 10^{-6}B_{T\rightarrow Xl\nu_l},\]
which is large as far as the semileptonic branching ratio $B_{T\rightarrow Xl\nu_l}$ measured for the top quark gets comparatively large, and much larger than $10^{-14}$, the SM prediction.

From the former analysis we can get 
\[{\cal B}r^{b^\prime}(t\rightarrow cZ)=\frac{4\pi}{\sin(2\theta)}{\cal B}r^{b^\prime}(t\rightarrow c\gamma)\approx 40 
{\cal B}r^{b^\prime}(t\rightarrow c\gamma),\]
two orders of magnitude larger than ${\cal B}r^{b^\prime}(t\rightarrow c\gamma)$, a value not far from the LHC capability, with a similar conclusion for the branching ${\cal B}r^{b^\prime}(t\rightarrow cg)$, where $g$ stands for the gluon field.

Finally we find 
\begin{eqnarray*}
{\cal B}r^{b^\prime}(t\rightarrow u\gamma)&\approx&\frac{|V_{ub^\prime}|^2}{|V_{cb^\prime}|^2}{\cal B}r^{b^\prime}(t\rightarrow c\gamma)\\
&\approx& 2.85\times 10^{-8} B_{T\rightarrow Xl\nu_l}.
\end{eqnarray*}

\subsection{\label{sec:sec56}Penguin processes for new quarks}
As can be seen from the former calculations, the GIM cancellation does not proceed for 3-3-1 models in general, mainly because the nonunitary character of $V^{4\times 5}_{mix}$, with the branching ratios proportional now to $F^Q(x)^2$, which is a function of $x=m^2_{q^\prime}/M_W^2\gg 1$, for $q^\prime=t^\prime,\; b^\prime,\; b^{\prime\prime}$.

To make predictions for the new quarks, a hierarchy between the heavy states must be assumed; for example, for $m_{t^\prime}>m_{b^\prime}\sim m_{b^{\prime\prime}}>m_t$, and scaling the branching ratio to the semileptonic decay $b^\prime\rightarrow Ul\nu_l$ for $U=t,c,u$, we get

\begin{equation}\label{finbpb}
\frac{{\cal B}r^{t}(b^\prime\rightarrow b\gamma)}{B_{B^\prime\rightarrow X_Ul\nu_l}}\approx\frac{3\alpha}{2\pi}
\frac{|V_{tb^\prime}^* V_{tb}F^{2/3}(x)|^2}{[f(x_t)|V_{tb^\prime}|^2]},
\end{equation}
which for $m_t=151$ GeV~\cite{koide} produces the result 
\[{\cal B}r^{t}(b^\prime\rightarrow b\gamma)\approx 2.4\times 10^{-4} B_{B^\prime\rightarrow X_Ul\nu_l}. \]
a value large enough to be detected at the LHC, even if the branching ratio $B_{T^\prime\rightarrow X_Bl\nu_l}$ is small.

\subsection{\label{sec:sec57}Meson mixing at tree-level}
The strongest constraint for the model under consideration here, comes from the new tree-level FCNC produced by the non-universal $Z^\prime$ neutral Gauge Boson. Ignoring CP-violating effects and using the results in Eq.~(\ref{Heff}) , the $K^{0}- \bar{K^{0}}$ mass difference produced by the physical $Z_2^\mu$ Gauge Boson, turns out to be
\begin{widetext}
\begin{equation}\label{dmz2}
(\Delta m_K)_{Z_2}=\frac{4\sqrt{2}G_FC^4_W C^2_\theta}{(3-4S^2_W)}|(V_5^d)_{32}^{*}(V_5^d)_{31}|^{2}\eta_{K}\left( \frac{M^{2}_{Z_1}}{M^{2}_{Z_2}}+T^{2}_\theta \right) B_Kf^{2}_Km_K,
\end{equation}
\end{widetext}
where the leading order QCD  corrections have been included through the parameter $\eta_{k}\approx 0.57$~\cite{Gilman}, $B_K$ and $f_K$ are the bag parameter and the decay constant for the kaon system respectively, and $C_\theta$ and $T_\theta$ are the cosine and tangent of the small mixing angle $\theta$ needed to define the physical fields $Z_1^\mu$ and $Z_2^\mu$.

As can be seen, for a small mixing angle $\theta,\;\Delta m_k$ is an inverse function of $M_{Z_2}^2$, the physical mass of the new neutral Gauge Boson. Our approach here is to use the experimental measured value $\Delta m_k$ to set a lower bound for $M_{Z_2}$.

Using the numerical values $G_F=1.166\times10^{-5}$ Gev$^{-2}$, $\theta_W=31.93°$, $M_{Z_1}=91.2$ Gev., $\Delta m_k=3.48\times 10^{-12}$ MeV., $\sqrt{B_K}f_K=135$ MeV, $m_k=497.65$ MeV; neglecting the small mixing angle $\theta$ and using $(V_5^d)_{32}^{*}(V_5^d)_{31}$ from the rotation matrix in (\ref{rot5}), the final value turns out to be $M_{Z_2}\geq 0.2$ TeV, one order of magnitude smaller than previous values calculated for this model~\cite{dwl}.

Now, for this down-up approach, there is no prediction coming from the $D^{0}- \bar{D^{0}}$ mixing (for which 
$\Delta m_D=4.607\times 10^{-11} $ MeV., $\sqrt{B_D}f_D=187$ MeV~\cite{Gilman}, $m_D=1864.5$ MeV, and $\eta_D\approx 0.57$) due to the zeroes in $V^u_4$. 

For the bottom sector we have for the $B^{0}_d- \bar{B}^0_d$ mixing, with $\Delta m_{B^0_d}=3.37\times 10^{-10}$ MeV., $\sqrt{B_B}f_B=208$ MeV~\cite{Gilman}, $m_B=5279.4$ MeV, and $\eta_B\approx 0.55$, that $M_{Z_2}\geq 2.1$ TeV. For the $B^{0}_s- \bar{B^{0}}_s$ mixing with $\Delta m_{B^0_s}=1.17\times 10^{-8}$ MeV., we obtain a limit $M_{Z_2}\geq 1.18$ TeV; both mass limits in agreement with the calculated value for this model, using precision measurements of the SM parameters~\cite{dwl}.

The conclusion here is that in general, for the down-up approach, the new neutral meson mixing, coming from the tree-level FCNC, do not violate current experimental measurements as far as 
\begin{equation}
M_{Z_2}\geq 2.1 {\rm TeV},
\end{equation}
mass value which justifies the assumption of neglecting the small mixing angle effects in Eq.~(\ref{dmz2})due to the fact that $T_\theta^2\leq 2.43\times 10^{-6}<<(M_{Z_1}/M_{Z_2})^2$.

But when the mixing angle is taken different from zero, there are new contributions to the meson mixing at tree-level, coming from the physical $Z_1^\mu$ Gauge Boson, given now by:
\begin{eqnarray}\nonumber
(\Delta m_K)_{Z_1}&=&
(\Delta m_K)_{Z_2}T_\theta^2\left[\frac{M^{2}_{Z_2}/M^{2}_{Z_1}+(C_\theta/S_\theta)^{2}}{M^{2}_{Z_1}/M^{2}_{Z_2}+T^{2}_\theta}\right] \\ \label{dmz1} &\leq& 0.3(\Delta m_K)_{Z_2},
\end{eqnarray}
where $S_\theta$ stands for the sine of the mixing angle $\theta$, and the numerical evaluation has been done for $M_{Z_2}\approx 2.1$ TeV, and $\theta^2=10^{-6}$.

\subsection{The up-down approach}
\begin{figure}
\includegraphics{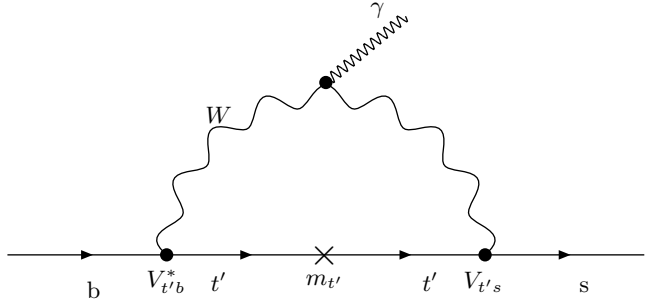}
\caption{\label{fig1}One-loop diagram contributing to the FCNC $b\longrightarrow s\gamma$}
\end{figure}

Next, let us quote the theoretical predictions for the up-down approach for which the rotation and mixing matrices in the appendix are used. In this approach, the mixing of the ordinary quarks with the exotic ones exists, but it is small due to the fact that $V_{tb}\sim 1$. Also, new penguin diagrams like the one depicted in Fig.~(\ref{fig1}) exist, due to the fact that for this approach $V_{t^\prime q}\neq 0$. The following is the list of our results:
\begin{eqnarray*}
{\cal B}r^{t^\prime}(b\rightarrow s\gamma)&\approx&\frac{3\alpha}{2\pi}
\frac{|V_{t^\prime b}^*V_{t^\prime s}F^Q(x^{t^\prime})|^2}{[f(x_c)|V_{cb}|^2+f(x_u)|V_{ub}|^2]}B_{B\rightarrow Xl\nu_l}\\
&\approx& 3.4\times 10^{-9}.
\end{eqnarray*}

\[{\cal B}r^{t^\prime}(b\rightarrow d\gamma)=\frac{|V_{t^\prime d}|^2}{|V_{t^\prime s}|^2}{\cal B}r^{t^\prime}(b\rightarrow s\gamma)\approx 3.6\times 10^{-10},\]

\begin{eqnarray*}
{\cal B}r^{t^\prime}(s\rightarrow d\gamma)&\approx&\frac{3\alpha}{2\pi}
\frac{|V_{t^\prime s}^*V_{t^\prime d}F^{2/3}(x^{t^\prime})|^2}{f(x^\prime_u)|V_{us}|^2}B_{K\rightarrow \pi l\nu_l}\\
&\approx&1.0\times 10^{-14}
\end{eqnarray*}
\begin{eqnarray*}
\frac{{\cal B}r^{b^\prime(b^{\prime\prime})}(c\rightarrow u\gamma)}{B_{D\rightarrow X_s l\nu_l}}&\approx&\frac{3\alpha}{2\pi}
\frac{|(V_{cb^\prime}^*V_{ub^\prime}+ V_{cb^{\prime\prime}}^*V_{ub^{\prime\prime}}) F^{-1/3}(x^{b^\prime})|^2}{[f(x_s)|V_{cs}|^2+f(x_d)|V_{cd}|^2]}\\
&\approx& \times 10^{-18}
\end{eqnarray*}

\begin{eqnarray*}
\frac{{\cal B}r^{b^\prime(b^{\prime\prime})}(t\rightarrow c\gamma)}{B_{T\rightarrow Xl\nu_l}}&\approx&\frac{3\alpha}{2\pi}
\frac{|(V_{tb^\prime}^*V_{cb^\prime}+V_{tb^{\prime\prime}}^*V_{cb^{\prime\prime}}) F^{-1/3}(x^{b^\prime})|^2}{[f(x_b)|V_{tb}|^2+f(x_s)|V_{ts}|^2]}\\
&\approx& 1.5\times 10^{-14}
\end{eqnarray*}
and finally 
\[\frac{{\cal B}r^{b^\prime(b^{\prime\prime})}(t\rightarrow u\gamma)}{B_T\rightarrow Xl\nu_l}\approx 2.3\times 10^{-15}.\]
All of them much smaller than the numbers calculated in the down-up approach, due to the now small mixing of the exotic quarks with the ordinary ones.

Recalculating the meson mixing processes for this up-down approach, the $M_{Z_2}$ mass value becomes now larger than 10 TeV in order to respect the experimental measurements (becomes larger than 12 TeV when the mixing is totally neglected, as it happens for example in the minimal 3-3-1 model of Pisano, Pleitez and Frampton~\cite{pf}).


\section {\label{sec:sec7} Conclusions}
The basic motivation of the present work was to study FCNC effects in the context of the 3-3-1 models with right-handed neutrinos. For this model there are four up-type quarks and five down-type quarks and its quark mixing matrix fails to be unitary. Besides, a new non-universal neutral current, able to produce FCNC effects at the tree level is present for this model.

For this analysis we searched for the largest mixing between ordinary and exotic quarks without violating current experimental constrains in the quark mixing matrix and in the values and bounds measured for FCNC processes.

Even though our analysis is ``ansatz" dependent, two main approaches, with different consequences, can be distinguish: the first one characterized by a value of $V_{tb}\sim 0.8$ and the second one for a value $V_{tb}\sim 1$. For the first approach the mixing of the ordinary quarks with the exotic ones is large, the penguin contributions to the FCNC are relevant and the tree-level meson mixing are perfectly under control for a mass $M_{Z_2}$ at the TeV scale. For the second approach the mixing of the ordinary quarks with the exotic ones is small, the penguin contribution to the FCNC are negligible, but the tree-level meson mixing became large, unless $M_{Z_2}$ gets a mass larger than 10 TeV.

The former conclusion is of relevance for the forthcoming Tevatron and LHC results, which should meassure with high accuracy the value of $V_{tb}$. In particular, a value of $V_{tb}\sim 1$ associated with a new non-universal neutral Gauge Boson below the TeV scale are almost incompatible, and in particular will rule out not only the 3-3-1 model with right-handed neutrinos, but also most of the 3-3-1 extensions of the SM. On the contrary, a value of $V_{tb}$ in the range $0.8\leq V_{tb}\leq 0.9$ can coexist with a new non-universal neutral Gauge Boson at the TeV scale, with strong predictions of rare top decays such as $t\rightarrow cZ$, with a branching ratio of the order of $10^{-5}$, perfectly reachable at the LHC~\cite{carv}.

FCNC produced by Higgs scalar Fields are not relevant for the 3-3-1 model with right-handed neutrinos. For the third family they do not exist at tree-level because the Higgs field $\phi_2$ which couples to the third family, does not couple to the other two families. For the first two families the processes may exist, but they are negligible small and proportional to $(m_sm_d/m_h^2)^2$ or to $(m_cm_u/m_h^2)^2$, where $m_h$ stands for the Higgs scalar mass.

Finally, let us mention that in the context of the 3-3-1 model with right-handed neutrinos, no FCNC effects at tree-level are present in the lepton sector, due to the universality for leptons present in the weak basis.

\vspace{1cm}

\appendix
\section{SM Textures}
In order to explain the known hierarchy of the quark masses and mixing angles, several ``ansatz" for up and down quark mass matrices have been suggested in the literature~\cite{hf}, some of them including the so-called texture zeros~\cite{SW}. In particular, symmetric mass matrices with four and five texture zeros were studied in detail in Refs.~\cite{HF, LI}, respectively. Unfortunately, precision measurements of several entries in the mixing matrix, rule out most of the suggested simple structures.

In this appendix we are going to introduce what we have called the up-down approach which consists in fitting the data (six quark masses and three mixing angles) to a unitary $3\times 3$ mixing matrix, and then allow this matrix to loose its unitary character by letting the ordinary quarks to mix with the exotic ones . Contrary to the approach used in the main text, this approach is characterized by the fact that $V_{tb}\sim 1$. Our numerical study suggest to start with the following hermitian, parallel, four texture zeros ansatz for the SM quark mass matrices 
\begin{equation}\label{uptex}
M^u_3 =hv\left(\begin{array}{ccc} 
0 & 0 & 11.4\lambda^4 \\ 
0 & 2.8\lambda^7 & 5.1\lambda^3  \\
11.4\lambda^4 & 5.1\lambda^3  & 1\\
\end{array}\right)=hv M^{0u}_3,
\end{equation}

\begin{equation}\label{dowtex}
M^d_3 =hv\left(\begin{array}{ccc} 
0 &   0 & 1.45\lambda^5+2i\lambda^7 \\ 
0  & -1.4\lambda^6 & 3\lambda^5+i\lambda^7  \\
1.45\lambda^5-2i\lambda^7 &  3\lambda^5-i\lambda^7  & 1.6\lambda^3 
\end{array}\right)=hv M^{0d}_3,
\end{equation}
where $h$ is a Yukawa coupling constants fixed by the top quark mass. The former ansatz for up and down quark mass matrices has the extra ingredient of being compatible with a new kind of flavor symmetry and its perturbative breaking as proposed by Froggatt and Nielsen~\cite{CD}, including a third order effect at the level of the bottom quark mass, implied by the entry $(M_3^{0d})_{33}=1.6\lambda^3$.

To check the validity of our ansatz let us use a value of $\lambda\approx 0.22$ and $hv=170$ GeV in matrices (\ref{uptex}) and (\ref{dowtex}) which produce the following quark mass values in units of MeV:
\[m_t=171500,\; m_c=614.4, \; m_u=2.3\]
\[m_b=2940,\; m_s=53.4, \; m_d=2.8;\]
numbers to be compared with the following values quoted from the second paper in Ref.~\cite{koide} (where they were calculates at the Fermi scale $\mu=M_Z$, using the $\overline{MS}$ scheme):
\begin{eqnarray}\nonumber
m_t&=&171700\pm 3000,\; m_c=619\pm 84, \; m_u=1.27^{+0.50}_{-0.42}\\ \label{mqex}
m_b&=&2890\pm 90,\; m_s=55^{+16}_{-15}, \; m_d=2.90^{+1.24}_{-1.19};
\end{eqnarray}

The rotation matrices which diagonalize the Hermitian mass matrices $M^u_3$ and $M^d_3$ in (\ref{uptex}) and (\ref{dowtex}) are given by
\begin{equation}\label{rotu}
V_3^u=\left(
\begin{array}{ccc}0.89397 & -0.44813 & 0.00046 \\-0.44735 & -0.89233 &0.06019 \\0.02656 & 0.05401 & 0.99819 \end{array}\right)_{rotu},
\end{equation}
and
\begin{equation}\label{rotd}
V_{3}^d=\left(
\begin{array}{ccc}0.97361 & 0.23347e^{-2.9i} & 0.02145e^{-3.8i}\\0.23043e^{2.9i} & 0.96825 & 0.09624e^{-0.92i}\\
0.04322e^{3.8i} & 0.08860e^{0.92i} & 0.99512 \end{array}
\right)_{rotd}.
\end{equation}

The consistency of our analysis shows up when we calculate the absolute values of $V_{CKM}=\sqrt{|V_3^uV_3^{d\dagger}|^2}$ which gives the following values
\begin{equation}\label{mix0}
V_{mix}^{(0)} =\left(\begin{array}{ccc} 
0.973 & 0.229 & -0.0033 \\ 
0.229 & -0.973 & 0.039 \\
0.0085 &  0.0377  & 0.999 \\
\end{array}\right),
\end{equation}
which is an (allmost) unitary matrix, in agreement with the experimental constrains quoted in matrix (\ref{maexp}).

Extending the previous analysis to the 3-3-1 model with right-handed neutrinos which includes four up type quarks and five down type quarks, we find that the maximall mixing allow of the ordinary quarks with the new ones, which does not violates the experimental values quoted in $V_{exp}$ in matrix (\ref{maexp}), neither the quark mass values quoted above, preserving the allmost unitary character of (\ref{mix0}), is given by

\begin{equation}\label{uptex4}
M^{u\prime}_4 =h_tv\left(\begin{array}{cccc} 
&& & 1.8\lambda^3 \\ 
&M^{0u}_{3x3}&&5\lambda^3  \\
&&&1\\
1.8\lambda^3 & 5\lambda^3 & 1 &10\\
\end{array}\right),
\end{equation}

\begin{equation}\label{dowtex5}
M^{d\prime}_5 =h_tv\left(\begin{array}{ccccc} 
&& & \lambda^6 & \lambda^4\\ 
&M^{0d}_{3x3}&&\lambda^5& \lambda^4  \\
&&& 0.6\lambda^2 & 2.5\lambda^2-i\lambda^4\\
\lambda^6& \lambda^5 & 0.6\lambda^2 & 10 & 1-i\lambda\\
\lambda^4& \lambda^4 & 2.5\lambda^2+i\lambda^4 &1 +i\lambda & 10 \\
\end{array}\right).
\end{equation}

The $4\times 4$ rotation matrix which diagonalize the Hermitian mass matrices $M^u_4$ in (\ref{uptex4}) is now given by
\begin{equation}\label{rot44}
V_4^{u\prime}=\left(
\begin{array}{cccc}-0.8936 & 0.4488 & 0.0002 & -0.0007 \\-0.4480 & -0.8919 & 0.0606 & -0.0005\\
-0.0273 & -0.0538 & -0.9922& 0.1093 \\
0.0022 & 0.0058 & 0.1092 & 0.9940 \end{array}\right)_{rotu},
\end{equation}
and the $5\times 5$ rotation matrix which diagonalize the Hermitian mass matrices $M^d_5$ in (\ref{dowtex5}) is now given by

\begin{widetext}
\begin{equation}\label{rotd5}
V_{5}^{d\prime}=\left(
\begin{array}{ccccc}0.9713 & 0.2368e^{-3i} & 0.0220e^{-4i} & 5.520\times 10{-5}e^{-4.7i}& 8.6\times 10{-4}e^{-1.7i}\\
0.2334e^{3i} & 0.9669 & 0.103e^{-0.96i} & 1.53\times 10^{-4}e^{-8.3i} & 9.51\times 10^{-4}e^{-2.5i}\\
0.0456e^{4i} & 0.095e^{0.96i} & 0.9944 & 1.72\times 10^{-3}e^{-8.4i} & 0.012 e^{-0.82i} \\
1.75\times 10^{-4}e^{-13.2i} & 1.46\times 10^{-4}e^{-16i} & 7.34e^{-14.8i} & 0.707 & 0.71e^{-12.4i} \\
1.6\times 10^{-4}e^{-0.57i} & 1.8\times 10^{-4}e^{-2.3i} & 9.6\times 10^{-3}e^{3.3i} & 0.706e^{12.4i} & 0.707
 \end{array}
\right)_{rotd}.
\end{equation}
\end{widetext}

Matrices that we combine as $V_{mix}^{4\times 5\prime}=\sqrt{|V_4^{u\prime}{\cal P}V_5^{d\prime\dagger}|^2}$, producing the following values
\begin{equation}\label{rotu4}
V_{mix}^{4\times 5\prime}=\left(
\begin{array}{ccccc}-0.9741 & 0.2260 & 0.0031 & 0.0001 & 0.0001 \\0.2260 & 0.9731 & 0.0449 & 0.0002 & 0.0003\\
0.0082 & 0.0439 & 0.9929 & 0.0073 & 0.0096 \\
0.0017 & 0.0051 & 0.1092 & 0.0008 & 0.0011 \end{array}\right)_{rotu}.
\end{equation}

To finish, let us mention that from our $3\times 3$ mass matrices (\ref{uptex}) and (\ref{dowtex}) we can obtain at the end a $V_{CKM}$ mixing matrix depending only of a single phase. As a matter of fact, we have chosen allready three arbitrary phases in the up quark sector such that the mass matrix $M^u$ becomes real. Then, two more phases can be eliminated from $V_3^d$ by a redefinition of the left-handed down quark fields, ending up with a single phase which propagates to $V_{CKM}=V_3^uV_3^{d\dagger}$. This single phase which shows up in a nonstandard parametrization of $V_{CKM}$ is the source of CP violation in the context of our ansatz.

\end{document}